\documentclass[onecolumn]{article}
\usepackage{amsfonts}
\usepackage{amsmath}
\usepackage{graphicx}

\input{tcilatex}

\begin{document}

\title{Generation and sudden death of entanglement in qubit-qutrit systems
in depolarizing noise}
\author{Salman Khan\thanks{%
sksafi@phys.qau.edu.pk} \\
Department of Physics, Quaid-i-Azam University, \\
Islamabad 45320, Pakistan}
\maketitle

\begin{abstract}
The dynamics of entanglement in some hybrid qubit-qutrit systems under the
influence of global, collective, local and multilocal depolarizing noise is
studied. It is shown that the depolarizing noise can be used to induce
entanglement. A critical point exists under every coupling of the system
with the environment at which all the states are equally entangled.
Furthermore, it is seen that no ESD occurs when either only the qubit is
coupled to its local environment or the system is coupled to multilocal or
global environment. This is an important result for various quantum
information processing tasks and hence needs further investigation.\newline

\end{abstract}

\section{Introduction}

Quantum entanglement is one of the potential sources of quantum theory. It
plays a vital role and works as a major resource for quantum communication
and computation \cite{springer}. Unveiling various aspects of quantum
entanglement has been a strenuous exercise for the last few decades. It is
frequently used in constructing many protocols, such as teleportation of
unknown states \cite{Bennett1}, quantum key distribution \cite{Ekert},
quantum cryptography \cite{Bennett2} and quantum computation \cite{Grover,
Vincenzo}. Completely isolating a quantum system from its environment is
rather an impossible task. It is, thus, necessary to investigate the
behavior of initial state entanglement in the presence of environmental
effects. The inevitable coupling between an environment and a quantum system
leads to the phenomenon of decoherence and it gives rise to an irreversible
transfer of information from the system to the environment \cite{Zurik,
Breuer, Zurik2}. Yu and Eberly \cite{Yu1, Yu2} showed that entanglement loss
occurs in a finite time under the action of pure vacuum noise in a bipartite
state of qubits. They found that, even though it takes infinite time to
complete decoherence locally, the global entanglement may be lost in finite
time. This phenomenon of sudden loss of entanglement has been named as
"entanglement sudden death" (ESD). The finite time loss of entanglement
definitely limits the application of entangled states in quantum information
processing. The phenomenon of ESD is not limited only to two qubit entangled
states, it is investigated in systems of larger spaces such as qutrits and
qudits \cite{Yonac, Jakobczyk, Ikram, Qasimi, Jeager1, Jeager2, Haung,
Jeager3}. A geometric interpretation of the effect of ESD is given in Ref. 
\cite{Terra}. The experimental evidences of the phenomenon of ESD have been
reported for optical setups \cite{Almeida} and atomic ensembles \cite{Laurat}%
.

Quantum states are grouped into separable and entangled states for
qubit-qubit and qubit-qutrit system by using Peres-Horodecki criterion \cite%
{Peres, Horodecki1} Nevertheless, such characterization for higher
dimensional bipartite states is difficult \cite{Horodecki2}. According to
this criterion, the partial transpose of a separable density matrix must has
non-negative eigenvalues. For a nonseparable state, the degree of
entanglement is quantified by the negativity, which is given by the sum of
the absolute values of the negative eigenvalues of the partial transpose of
the density matrix.

In this paper I study the behavior of entanglement of a hybrid qubit-qutrit
system in the presence of depolarizing noise. The individual qubit and
qutrit states of the system, I consider, are incoherent but the composite
system may still possess coherence and entanglement. For detailed study of
the properties of such systems, I refer the readers to Ref. \cite{Ann},
where the authors showed the existence of ESD in such composite systems
under the influence of dephasing noise. I consider various couplings of the
system and environment in which the system is influenced by global,
collective, local, or multilocal depolarizing noise. I show that although
ESD occurs only in certain states under particular situations for various
coupling of the system and environment, however, it can be controlled and
completely avoided if certain measurement are taken. For example, under the
action of multilocal and global noise, all the states avoid ESD if one or
the other environment is controlled. Furthermore, I show that the degree of
entanglement regrows for all states in the range of large values of
decoherence parameters.

\subsection{Qubit-Qutrit System in Depolarizing Noise}

I consider a composite system of a qubit A and a qutrit B that are coupled
to a noisy environment both collectively and individually. The collective
coupling refers to the situation when both the qubit and qutrit are
influenced by the same environment, whereas the local and multilocal
couplings describe the situations when either the qubit or qutrit or both
qubit and qutrit are independently influenced by its own environment. The
system is said to be coupled to a global environment when it is influenced
by both collective and multilocal noises at the same time. Let the bases of
Hilbert space of the qubit be denoted by $|0\rangle $ and $|1\rangle $ and
that of the qutrit by $|0\rangle $, $|1\rangle $ and $|2\rangle $. Then the
bases of the composite system are given in the order $|00\rangle $, $%
|01\rangle $, $|02\rangle $, $|10\rangle $, $|11\rangle $, $|12\rangle $.

The dynamics of the composite system in the presence of depolarizing noise
can best be described in the Kraus operators formalism. The Kraus operators
for a single qubit depolarizing noise are given as%
\begin{eqnarray}
E_{o}^{A} &=&\sqrt{1-p}\sigma _{0},\qquad E_{1}^{A}=\sqrt{p/3}\sigma _{1}, 
\notag \\
E_{2}^{A} &=&\sqrt{p/3}\sigma _{2},\qquad E_{3}^{A}=\sqrt{p/3}\sigma _{3},
\label{E1}
\end{eqnarray}%
where $\sigma _{i}$ are the Pauli matrices. The Kraus operators for a single
qutrit depolarizing noise are given as \cite{Salimi}%
\begin{eqnarray}
E_{0}^{B} &=&\sqrt{1-p}I_{3},\quad E_{1}^{B}=\sqrt{\frac{p}{8}}Y,\quad
E_{2}^{B}=\sqrt{\frac{p}{8}}Z,  \notag \\
E_{3}^{B} &=&\sqrt{\frac{p}{8}}Y^{2},\quad E_{4}^{B}=\sqrt{\frac{p}{8}}%
YZ,\quad E_{5}^{B}=\sqrt{\frac{p}{8}}Y^{2}Z,  \notag \\
E_{6}^{B} &=&\sqrt{\frac{p}{8}}YZ^{2},\quad E_{7}^{B}=\sqrt{\frac{p}{8}}%
Y^{2}Z^{2},\quad E_{8}^{B}=\sqrt{\frac{p}{8}}Z^{2},  \label{1}
\end{eqnarray}%
with

\begin{equation}
Y=\left( 
\begin{array}{ccc}
0 & 1 & 0 \\ 
0 & 0 & 1 \\ 
1 & 0 & 0%
\end{array}%
\right) ,\quad Z=\left( 
\begin{array}{ccc}
1 & 0 & 0 \\ 
0 & \omega & 0 \\ 
0 & 0 & \omega ^{2}%
\end{array}%
\right) ,  \label{2}
\end{equation}%
where $\omega =e^{2i\pi /3}$ and $I_{3}$ is the identity matrix of order $3$%
. In Eqs. (\ref{E1}) and (\ref{1}) $p=\left[ 0,1\right] $ is the decoherence
parameter. The lower and upper limits of $p$ stand, respectively, for
undecohered and fully decohered cases of the noisy environment. The Kraus
operators for both qubit and qutrit satisfy the completeness relation $%
\sum_{i}E_{i}^{\dag }E_{i}=I$. The evolution of the initial density matrix
of the system when it is influenced by the global depolarizing noise is
given in the Kraus operators formalism as follow%
\begin{equation}
\rho ^{\prime }=\sum_{i,j,k}\left( E_{i}^{AB}E_{j}^{B}E_{k}^{A}\right) \rho
\left( E_{k}^{A\dag }E_{j}^{B\dag }E_{i}^{AB\dag }\right) ,  \label{3}
\end{equation}%
where $E_{k}^{A}=E_{m}^{A}\otimes I_{3}$, $E_{j}^{B}=\sigma _{0}\otimes
E_{n}^{B}$ are the Kraus operators of the multilocal coupling of the qubit
and the qutrit individually and $E_{i}^{AB}$ are the Kraus operators of the
collective coupling that are formed from the tensor product of the Kraus
operators of a single qubit and a single qutrit depolarizing noise in the
form $E_{m}^{A}\otimes E_{n}^{B}$. The subscripts $m=0,1,2,3$, and $%
n=0,1,2,...8$ stand, respectively, for a single qubit and a single qutrit
Kraus operators of a depolarizing noise. The initial density matrix $\rho $
of the composite system is given by the following one parameter family of
matrices%
\begin{equation}
\rho (x)=\left( 
\begin{array}{cccccc}
\frac{1}{4} & 0 & 0 & 0 & 0 & x \\ 
0 & \frac{1}{8} & 0 & 0 & 0 & 0 \\ 
0 & 0 & \frac{1}{8} & 0 & 0 & 0 \\ 
0 & 0 & 0 & \frac{1}{8} & 0 & 0 \\ 
0 & 0 & 0 & 0 & \frac{1}{8} & 0 \\ 
x & 0 & 0 & 0 & 0 & \frac{1}{4}%
\end{array}%
\right) ,  \label{4}
\end{equation}%
where $0\leq x\leq \frac{1}{4}$. For this range of values of $x$, all the
eigenvalues are positive and the matrix is a well-defined density matrix.
The partial transpose criterion for this density matrix shows that in the
specified range of $x$, it is entangled (see Ref. \cite{Ann}). Using initial
density matrix of Eq. (\ref{4}) in Eq. (\ref{3}) and taking the partial
transpose over the qubit, it is easy and straightforward to find the
eigenvalues. Let the decoherence parameters for local noise of the qubit and
qutrit and collective noise of the composite system be $p_{1}$, $p_{2}$ and $%
p$ respectively. Then, the eigenvalues of the partial transpose of the final
density matrix when only the qubit is coupled to the noisy environment are
given by%
\begin{eqnarray}
\lambda _{1,2} &=&\frac{1}{4}-\frac{3}{32}p_{1},  \notag \\
\lambda _{3,4} &=&\frac{1}{8}+\frac{3}{64}p_{1},  \notag \\
\lambda _{5} &=&\frac{1}{8}+\frac{3}{64}\left( 1-24x\right) p_{1}+x  \notag
\\
\lambda _{6} &=&\frac{1}{8}+\frac{3}{64}\left( 1+24x\right) p_{1}-x.
\label{5}
\end{eqnarray}%
\begin{figure}[h]
\begin{center}
\begin{tabular}{ccc}
\vspace{-0.5cm} \includegraphics[scale=1.2]{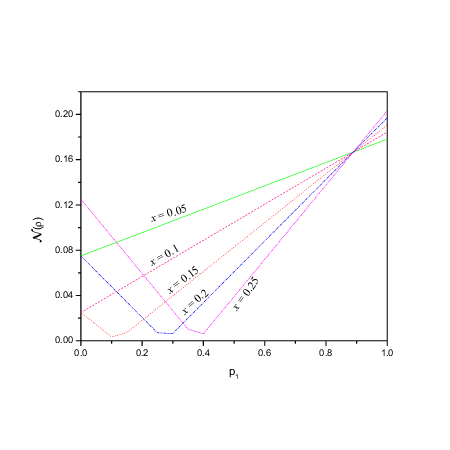}\put(-320,220) &  & 
\end{tabular}%
\end{center}
\caption{The negativity for the case when only the qubit is coupled to the
environment is plotted for different density matrices against the
decoherence parameter $p_{1}$.}
\label{Figure1}
\end{figure}
The only possible negative eigenvalue is the last one. The degree of
entanglement according to the definition of negativity becomes%
\begin{equation}
\mathcal{N}(\rho )=\max \left\{ 0,\left\vert \frac{1}{8}+\frac{3}{64}\left(
1+24x\right) p_{1}-x\right\vert \right\} ,  \label{E5}
\end{equation}%
To observe the dynamics of entanglement, I plot the negativity for various
values of $x$, which represents different density matrices, against the
decoherence parameter $p_{1}$ in Fig. 1. The figure shows that the behavior
of negativity splits the density matrices under consideration into two
groups. In one case, the negativity increases linearly with increasing
values of $p_{1}$. In the other case, the negativity first decreases to a
minimum, but positive, for a particular value of $p_{1}$ and then increases
linearly as $p_{1}$ increases. However no ESD occurs for the whole range of
decoherence parameter for any density matrix. There is a critical point at
which, irrespective of the values of $x$, all the density matrices reach to
the same degree of entanglement and it happens at $p_{1}=0.888$. Beyond this
value of $p_{1}$, the degree of entanglement is higher for density matrices
that correspond to large values of $x$. 
\begin{figure}[h]
\begin{center}
\begin{tabular}{ccc}
\vspace{-0.5cm} \includegraphics[scale=1.2]{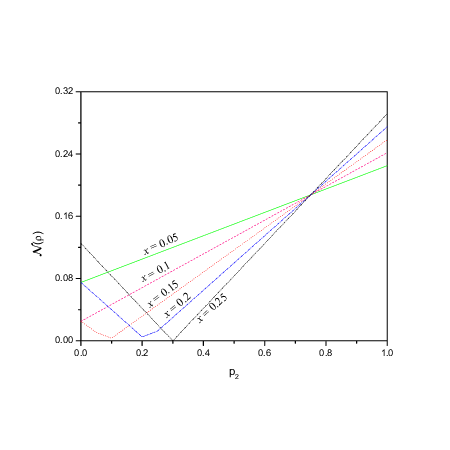}\put(-320,220) &  & 
\end{tabular}%
\end{center}
\caption{The negativity for the case when only the qutrit is coupled to the
environment is plotted against the decoherence parameter $p_{2}$ for
different values of the parameter $x$.}
\label{Figure2}
\end{figure}

The eigenvalues of the partial transpose matrix when only the qutrit is
coupled to local depolarizing noise are given by%
\begin{eqnarray}
\lambda _{1,2} &=&\frac{1}{4}-\frac{3}{32}p_{1}-\frac{1}{12}p_{2}+\frac{3}{32%
}p_{1}p_{2}  \notag \\
\lambda _{3,4} &=&\frac{1}{8}+\frac{3}{64}p_{1},  \notag \\
\lambda _{5} &=&\frac{1}{8}+\frac{1}{12}\left( 1-16x\right) p_{2}+x,  \notag
\\
\lambda _{6} &=&\frac{1}{8}+\frac{1}{12}\left( 1+16x\right) p_{2}-x.
\label{6}
\end{eqnarray}%
The only eigenvalue that possibly becomes negative is the last one. The
negativity in this case becomes%
\begin{equation}
\mathcal{N}(\rho )=\max \left\{ 0,\left\vert \frac{1}{8}+\frac{1}{12}\left(
1+16x\right) p_{2}-x\right\vert \right\} .  \label{E6}
\end{equation}%
The behavior of entanglement in this case is nearly similar to the case of
only qubit undergoing decoherence. Besides the rate of change in variation
of negativity, ESD does occurs in the range of large values of $x$ and the
critical point at which all the density matrices are equally entangled
shifts to $p_{2}=0.75$. Moreover, the degree of entanglement for large
values of $x$ for a fully decohered case is higher as compared to the case
of only qubit decoherence. The negativity in this case is plotted against
the decoherence parameter $p_{2}$ in Fig. 2. 
\begin{figure}[h]
\begin{center}
\begin{tabular}{ccc}
\vspace{-0.5cm} \includegraphics[scale=1.2]{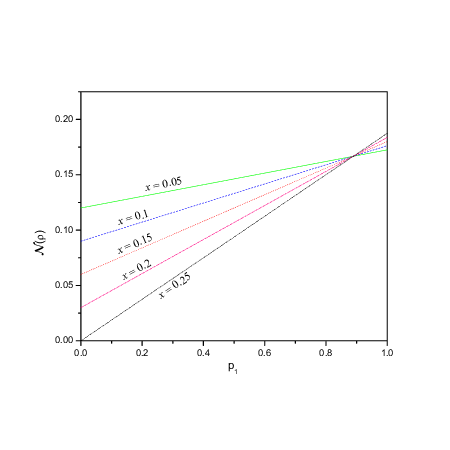}\put(-320,220) &  & 
\end{tabular}%
\end{center}
\caption{The negativity for the case of multilocal noise is plotted against
the decoherence parameter $p_{1}$ for various values of the parameter $x$
for $p_{2}=0.3$.}
\label{Figure3}
\end{figure}

The eigenvalues of the partial transpose of the final density matrix for the
multilocal case become

\begin{eqnarray}
\lambda _{1,2} &=&\frac{1}{12}\left( 3-p_{2}\right) ,  \notag \\
\lambda _{3,4} &=&\frac{1}{8},  \notag \\
\lambda _{5} &=&\frac{1}{8}+\frac{3}{64}p_{1}+\frac{1}{12}p_{2}-\frac{3}{32}%
\left( 1-16x\right) p_{1}p_{2}  \notag \\
&&-\left( \frac{9}{8}p_{1}+\frac{4}{3}p_{2}-1\right) x,  \notag \\
\lambda _{6} &=&\frac{1}{8}+\frac{3}{64}p_{1}+\frac{1}{12}p_{2}-\frac{3}{32}%
\left( 1+16x\right) p_{1}p_{2}  \notag \\
&&+\left( \frac{9}{8}p_{1}+\frac{4}{3}p_{2}-1\right) x.  \label{E7}
\end{eqnarray}%
Again, the only one possibly negative eigenvalue is the last one. The
negativity in this case can be written as%
\begin{equation}
\mathcal{N}(\rho )=\max \left\{ 0,\left\vert 
\begin{array}{c}
\frac{1}{8}+\frac{3}{64}p_{1}+\frac{1}{12}p_{2}-\frac{3}{32}\left(
1+16x\right) p_{1}p_{2} \\ 
+\left( \frac{9}{8}p_{1}+\frac{4}{3}p_{2}-1\right) x.%
\end{array}%
\right\vert \right\} .  \label{7}
\end{equation}%
The negativity for the case of multilocal noise is plotted in Fig. $3$
against the decoherence parameter $p_{1}$ for $p_{2}=0.3$. It can be seen
from the figure that no density matrix in the chosen range of $x$, undergoes
ESD. It is true for $0.3\leq p_{2}<1$. For $p_{2}<0.3$, the ESD occurs not
for all but only for certain density matrices in the upper limit of $x$. The
degree of entanglement increases for density matrices of large $x$ as $p_{1}$
increases when $0.32\leq p_{2}\leq 0.6538$. Whereas for $0.6538\leq p_{2}<1$%
, the degree of entanglement decreases with increasing values of $p_{1}$.
This is shown in Fig. $4$. A nearly similar behavior of the negativity is
observed when $p_{1}$ is kept constant. 
\begin{figure}[h]
\begin{center}
\begin{tabular}{ccc}
\vspace{-0.5cm} \includegraphics[scale=1.2]{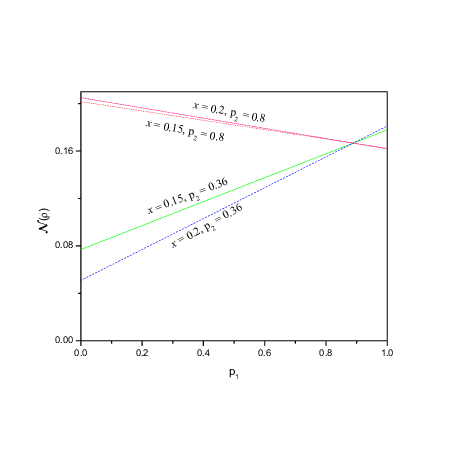}\put(-320,220) &  & 
\end{tabular}%
\end{center}
\caption{The negativity for the case of multilocal noise is plotted against
the decoherence parameter $p_{1}$ for different set of values of decoherence
parameter $p_{2}$ and the parameter $x$.}
\label{Figure4}
\end{figure}

For collective noise the eigenvalues are obtained by replacing $p_{1}=p_{2}$
in Eq. (\ref{E7}). In this case the negativity becomes%
\begin{equation}
\mathcal{N}(\rho )=\max \left\{ 0,\left\vert \frac{1}{8}+\frac{1}{32}\left( 
\frac{25}{6}-3p\right) p\left( \frac{59}{24}-\frac{3}{2}p\right)
px-x\right\vert \right\} .  \label{8}
\end{equation}%
The effect of decoherence on the negativity is shown in Fig. $5$. The
behavior of negativity in this case is nearly similar as discussed
previously. However, the ESD occurs only in the intermediate range of values
of $x$. Unlike the previous cases, there are two critical points that
happens at $p=0.75,0.888$. Beyond the first critical point the negativity
for lower values of $x$ decreases and for higher values of $x$, it first
increases and then drops, reaching the second critical point as $p$
increases. 
\begin{figure}[h]
\begin{center}
\begin{tabular}{ccc}
\vspace{-0.5cm} \includegraphics[scale=1.2]{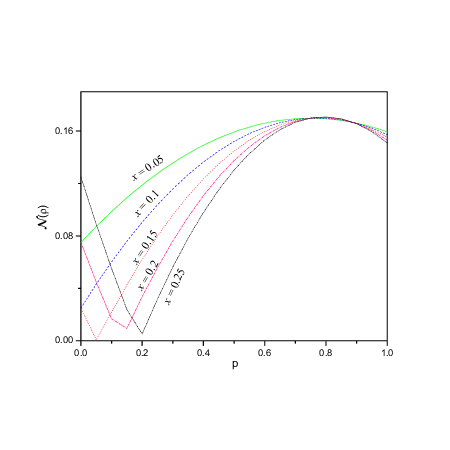}\put(-320,220) &  & 
\end{tabular}%
\end{center}
\caption{The negativity for the case of collective noise is plotted against
the decoherence parameter $p$ for various value of the parameter $x$.}
\label{Figure5}
\end{figure}

Finally, I consider the influence of global depolarizing noise. The general
form of the eigenvalues of the partial transpose of the final density matrix
for this case have quite lengthy expressions. Instead of writing the general
expressions for all of them, I consider only the one that can possibly
become negative, which is the last one as before. For a special case in
which the multilocal decoherence parameters $p_{1}=p_{2}=1/2$, it becomes%
\begin{equation}
\lambda _{6}=\frac{1}{4608}[96(8-7x)-63p^{2}(1+16x)+28p(2+59x)]  \label{9}
\end{equation}%
The corresponding negativity becomes%
\begin{equation}
\mathcal{N}(\rho )=\max \left\{ 0,\left\vert \frac{%
[2184-4800x-450p^{2}(1+16x)+5p(107+2360x)]}{13824}\right\vert \right\} .
\label{10}
\end{equation}%
\begin{figure}[h]
\begin{center}
\begin{tabular}{ccc}
\vspace{-0.5cm} \includegraphics[scale=1.2]{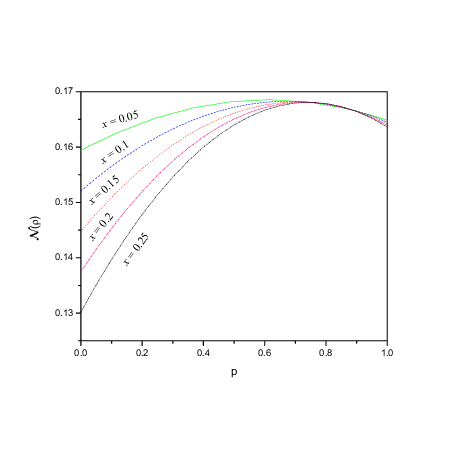}\put(-320,220) &  & 
\end{tabular}%
\end{center}
\caption{The negativity for the case of global noise when $p_{1}=p_{2}=0.5$
is plotted against the decoherence parameter $p$ for various values of the
parameter $x$.}
\label{Figure6}
\end{figure}
The behavior of negativity as a function of $p$ for various values of $x$ is
shown in Fig. $6$. It is clear from the figure that ESD can be completely
avoided under the action of global environment for the chosen values of
multilocal decoherence parameters. However, this is not true for all values
of multilocal decoherence parameters. For example, for $p_{1}=p_{2}=1/10$,
some density matrices in the range of large values of $x$ exhibit ESD.
Furthermore, for $p_{1}=p_{2}=3/4$, the negativity becomes independent of $x$
and all the density matrices respond equally to $p$.

\section{Summary}

I study the dynamics of entanglement for certain hybrid qubit-qutrit states
under global, collective, local and multilocal depolarizing noise. I show
that unlike the case of dephasing noise \cite{Ann}, the influence of
depolarizing noise is different for different coupling of the system and the
environment. Using partial transpose criterion for quantifying entanglement,
it is shown that not all but only certain initially entangled density
matrices exhibit ESD in depolarizing noise in the range of lower values of
decoherence parameter under particular coupling. No ESD in any density
matrix occurs when only the qubit is coupled to its local environment. The
only density matrices that corresponds to higher values of $x$ exhibit ESD
when only the qutrit is coupled to its local environment. The conjecture
made in Ref. \cite{Ann} that ESD is a generic phenomenon to occur in all
bipartite quantum system is not correct. For every density matrix that
undergoes ESD at a particular value of decoherence parameter, the re-birth
of entanglement occurs at values higher than that particular value of
decoherence parameter. The degree of entanglement increases for large values
of decoherence parameter for all states under every possible coupling of the
system and the environment considered here. A critical point, at which all
the states has equal degree of entanglement, is observed. It is shown that
in the case of only qutrit-environment's coupling the decrease in negativity
occurs faster in certain density matrices as compared to only the
qubit-environment's coupling. The entanglement sudden death happens only in
the intermediate values of $x$ under the action of collective noise.
Furthermore, it is shown that the ESD can be completely avoided under
certain situations when the system is under the action of multilocal and
global noise.\newline

{\LARGE Figure Captions}\newline
Figure $1$. The negativity for the case when only the qubit is coupled to
the environment is plotted for different density matrices against the
decoherence parameter $p_{1}$.\newline
Figure $2$. The negativity for the case when only the qutrit is coupled to
the environment is plotted against the decoherence parameter $p_{2}$ for
different values of the parameter $x$.\newline
Figure $3$. The negativity for the case of multilocal noise is plotted
against the decoherence parameter $p_{1}$ for various values of the
parameter $x$ for $p_{2}=0.3$.\newline
Figure $4$. The negativity for the case of multilocal noise is plotted
against the decoherence parameter $p_{1}$ for different set of values of
decoherence parameter $p_{2}$ and the parameter $x$.\newline
Figure $5$. The negativity for the case of collective noise is plotted
against the decoherence parameter $p$ for various value of the parameter $x$.%
\newline
Figure $6$. The negativity for the case of global noise when $%
p_{1}=p_{2}=0.5 $ is plotted against the decoherence parameter $p$ for
various values of the parameter $x$.\newline

\end{document}